# Indication of unconventional superconductivity in layered bismuth chalcogenide superconductor $LaO_{0.6}F_{0.4}Bi(S,Se)_2$ revealed by Se isotope effect

Kazuhisa Hoshi, Yosuke Goto, Yoshikazu Mizuguchi*

Department of Physics, Tokyo Metropolitan University, 1-1, Minami-osawa, Hachioji 192-0397, Japan.

Abstract

We have investigated the Se isotope effect in layered bismuth chalcogenide ($BiCh_2$-based) superconductor $LaO_{0.6}F_{0.4}Bi(S,Se)_2$ with $^{76}Se$ and $^{80}Se$. For all examined samples, the Se concentration, which is linked to the superconducting properties, is successfully controlled within $x = 1.09–1.14$ in $LaO_{0.6}F_{0.4}BiS_{2-x}Se_x$. From the magnetization and electrical resistivity measurements, changes in $T_c$ are not observed for the $LaO_{0.6}F_{0.4}Bi(S,Se)_2$ samples with $^{76}Se$ and $^{80}Se$ isotopes. Our results suggest that pairing in the $BiCh_2$-based superconductors is not mediated by phonons, and unconventional superconductivity states may emerge in the $BiCh_2$ layers of $LaO_{0.6}F_{0.4}Bi(S,Se)_2$.

In most superconductors, electron–phonon interactions are responsible for the emergence of superconductivity [1]. According to the Bardeen–Cooper–Schrieffer (BCS) theory [1], the transition temperature ($T_c$) of a conventional superconductor is proportional to its phonon energy $\hbar\omega$, where $\hbar$ and $\omega$ are the Planck constant and phonon frequency, respectively. Therefore, the $T_c$ of conventional electron–phonon superconductors is sensitive to the isotope effect of the constituent elements. The isotope exponent $\alpha$ is defined by $T_c \sim M^{-\alpha}$, where $M$ is the isotope mass, and $\alpha \sim 0.5$ is expected from the BCS theory [1]. Therefore, the isotope effect has been used to investigate whether the mechanisms of superconductors are conventional or unconventional. Indeed, $\alpha$ close to 0.5 has been reported in $(Ba,K)BiO_3$ ($\alpha_O \sim 0.5$) [2], doped fullerene ($\alpha_C \sim 0.4$) [3], $MgB_2$ ($\alpha_B \sim 0.3$) [4], and Ni- and Pd-based borocarbides ($\alpha_B \sim 0.3$) [5]. In superconductors with unconventional mechanisms, the isotope effect is not consistent with the expectation from the BCS theory. In the cuprate superconductor system, $\alpha$ deviates from 0.5 and shows anomalous dependence on carrier concentration [6,7]. In the Fe-based superconductor, one research group reported $\alpha_{Fe} \sim 0.4$ in $SmFeAs(O,F)$ and $(Ba,K)Fe_2As_2$ [8], but another group reported an inverse isotope effect, negative $\alpha_{Fe}$, for the same composition [9].

In 2012, a new layered superconductor system with a $BiS_2$ superconducting layer was discovered [10,11]. Since the crystal structure of $BiS_2$-based superconductors resembled those of cuprate and Fe-based superconductors, researchers have explored new $BiS_2$-based superconductors with higher $T_c$. Six typical superconductor systems with $BiCh_2$-type (Ch: S, Se) superconducting layers and various kinds of blocking layers have been discovered, and the highest record of $T_c$ is 11 K [12,13]. The pairing mechanisms of superconductivity in the $BiCh_2$-based system have been unexplained [14]. In early theoretical calculations, conventional phonon-mediated pairing mechanisms were proposed [15]. In addition, Raman scattering, muon-spin spectroscopy measurements ($\mu$SR), and thermal conductivity experiments suggested conventional mechanisms with a fully gapped s-wave [16-18]. However, recent theoretical calculations indicated that $T_c$ with the order of several to 10 K in the $BiS_2$-based superconductor cannot be explained within existing conventions [19]. Furthermore, angle-resolved photoemission spectroscopy (ARPES) proposed unconventional pairing mechanisms owing to the observation of the highly anisotropic superconducting gap in $NdO_{0.71}F_{0.29}BiS_2$ [20]. Therefore, we have demonstrated the isotope effect in a $BiCh_2$-based system $La(O,F)Bi(S,Se)_2$.

On the target phase of this study, our recent studies on $LaO_{0.6}F_{0.4}Bi(S,Se)_2$ revealed that the emerging superconducting states in this Se-substituted system are quite homogeneous owing to the enhanced in-plane chemical pressure effect and the suppressed local in-plane disorder [21-23]; local in-plane disorder can intrinsically exist in the $BiCh_2$-based system owing to Bi lone pair electrons [24]. In addition, $T_c$ in the $LaO_{1-x}F_xBiSSe$ does not change for $x = 0.2$–0.5, indicating that $T_c$ is insensitive to electron doping amount in this region. Among those, the sample with $x = 0.4$ shows the sharpest superconducting transition [23]. Based on these facts, we choose the nominal composition of LaO$_{0.6}$F$_{0.4}$BiSSe to use in the isotope experiments. Since there is no stable isotope of Bi other than $^{209}$Bi, and Bi is too heavy for the precise investigation of the isotope effect, we used $^{76}$Se and $^{80}$Se isotopes. The superconductivity in BiCh$_2$-based systems emerges in the BiCh plane, and the conduction band is composed of Bi-6p orbitals hybridized with Ch-p orbitals [14,25]. In the present system, Se selectively occupies the in-plane site [23]. Therefore, the lattice vibration of Bi and Se should be responsible for the superconductivity, if the conventional phonon-mediated mechanisms are working. The Se isotope effect should then be observed in LaO$_{0.6}$F$_{0.4}$BiSSe if phonon-mediated. Assuming $\alpha_{Se} = 0.5$ ($\alpha$ expected from BCS theory) and $T_c \sim 3.8$ K [23], the difference in $T_c$ ($\Delta T_c$) is expected to be 0.098 K between the samples with $^{76}$Se and $^{80}$Se. When $\alpha_{Se} = 0.2$ and 0.3, which is close to the values observed in MgB$_2$ and borocarbide, $\Delta T_c$ is expected to be 0.039 and 0.059 K, respectively. Therefore, before synthesizing isotope samples, we optimized the synthesis procedure of the polycrystalline LaO$_{0.6}$F$_{0.4}$BiSSe samples using conventional Se powders to investigate very small $\Delta T_c$ between $^{76}$Se and $^{80}$Se.

As a result of the isotope-effect study, we found that the changes in $T_c$ between the $^{76}$Se and $^{80}$Se samples, estimated from magnetization and electrical resistivity measurements, are apparently smaller than that expected from phonon-mediated mechanisms. Our results suggest that pairing in the BiCh$_2$-based superconductors is not mediated by phonons, and unconventional superconductivity may emerge in the BiCh$_2$ layer.

Polycrystalline samples with a starting nominal composition of LaO$_{0.6}$F$_{0.4}$BiSSe were prepared by a solid-state-reaction method. Powders of La$_2$S$_3$ (99.9%), La$_2$O$_3$ (99.9%), Bi$_2$O$_3$ (99.99%), BiF$_3$ (99.9%), and Bi (99.999%) gains were used. Powders of Se isotopes $^{76}$Se (99.80%) and $^{80}$Se (99.91%) were purchased from ISOFLEX. The powders with the starting nominal composition of LaO$_{0.6}$F$_{0.4}$BiSSe were mixed using a mortar, pressed into pellets, sealed into an evacuated quartz tube, and heated at 700 °C for 20 h. The product was ground, mixed for homogenization, pressed into pellets, and annealed in an evacuated quartz tube at 700 °C for 20 h. X-ray diffraction patterns were collected by a Rigaku X-ray diffractometer with Cu-K$\alpha$ radiation using the $2\theta$-$\theta$ method with a range of $2\theta = 10$–120°. The obtained X-ray patterns were refined using the Rietveld method [26]. To obtain better refinement, a secondary phase of BiF$_3$, typically 3% mass fraction, was included in the refinements. In addition, very small impurity peaks, possibly LaS, were observed in the diffraction patterns. In the Rietveld refinements, occupancy at the O/F site was fixed at O$_{0.6}$F$_{0.4}$ because O and F cannot be reliably refined using X-ray diffraction data. Isotropic displacement parameters were fixed as the values obtained from the refinement of the synchrotron XRD data [23]. A schematic image of the crystal structure was captured using VESTA [27]. The temperature dependence of magnetization was measured after both zero-field cooling (ZFC) and field cooling (FC) using a superconducting quantum interference devise (SQUID) magnetometer with an applied field of 20 Oe by a Magnetic Property Measurement System (MPMS-3). The onset $T_c$ in the magnetization measurements was defined using a cross point of two linear fitting lines as shown in Fig. 2b–e. The irreversible temperature $T_{irr}$ was defined as the temperature at which the difference between ZFC and FC curves emerged. The temperature dependence of electrical resistivity was measured using a four-terminal method with a current of 1 mA by a Physical Property Measurement System (PPMS, Quantum Design). The $T_c$ in the electrical resistivity measurements was defined as the temperature where zero-resistivity state was observed. To precisely discuss the isotope effect, we have compared the resistivity transitions for three samples measured together on the same sample pack.

As seen in Fig. 1, using Rietveld refinement of the X-ray diffraction pattern, we confirmed that the obtained samples were structurally comparable in regard to impurity amount, lattice constant, and chalcogen (S/Se) concentration, as listed in Table 1. After Rietveld refinements, we found that the actual Se concentration was slightly larger than that of S in the obtained samples. Particularly, Se concentration ($x$) affects the superconducting properties in $LaO_{0.6}F_{0.4}BiS_{2-x}Se_x$; hence, we compare the superconducting properties of $x$ = 1.09–1.14 only, to precisely discuss the isotope effects. The detailed information about crystal structure parameters is summarized in Table 1. Here, we show the superconducting properties of two samples with $^{76}$Se, labelled as $^{76}$Se–#1 and $^{76}$Se–#2, and two samples with $^{80}$Se, labelled as $^{80}$Se–#1 and $^{80}$Se–#2.

Figure 2(a) and 2(b) show the temperature dependences of magnetization for the $^{76}$Se–#1, $^{76}$Se–#2, $^{80}$Se–#1, and $^{80}$Se–#2 samples. For all samples, a sharp transition and a large diamagnetic signal are observed. The enlarged figures of the magnetization around the superconducting transition are displayed in Fig. 2(c–f). Surprisingly, the onset $T_c$ does not change within 0.01 K for all samples. The onset $T_c$ is estimated as 3.77, 3.76, 3.76, and 3.77 K for $^{76}$Se–#1, $^{76}$Se–#2, $^{80}$Se–#1, and $^{80}$Se–#2, respectively. In addition, irreversible temperature $T_{irr}$, defined as the temperature at which the difference between ZFC and FC curves emerged and corresponding to the emergence of the superconducting current path, is estimated to be the same, $T_{irr}$ = 3.74 K. Figure 3 shows the temperature dependences of normalized electrical resistivity [$\rho(T) / \rho(4\ K)$] for $^{76}$Se–#1, $^{80}$Se–#1 and $^{80}$Se–#2, which were measured together on the same sample pack. The resistivity data was normalized using resistivity at 4 K for comparison of $T_c$. The $T_c$ is estimated as 3.73, 3.73, and 3.72 K for $^{76}$Se–#1, $^{80}$Se–#1 and $^{80}$Se–#2, respectively.

Based on the observed $\Delta T_c$ in the magnetization and electrical resistivity, we conclude that $\alpha_{Se}$ in $LaO_{0.6}F_{0.4}BiS_{2-x}Se_x$ with $x$ = 1.09–1.14 is very close to zero. According to our Rietveld analyses (see Table 1), the Ch1 site (in-plane site) is almost completely occupied with Se: Se occupancy at the Ch1 site is 95–99%. Therefore, phonons, at least in-plane phonons, should not be responsible for pairing in the superconductivity of $LaO_{0.6}F_{0.4}Bi(S,Se)_2$. This conclusion is consistent with the theoretical calculations by Morice *et al.* [19]. Although the pairing mechanisms for the superconductivity of $LaO_{0.6}F_{0.4}Bi(S,Se)_2$ cannot be completely clarified with the present isotope effect only, we briefly discuss the possibility of unconventional superconductivity. As mentioned above, unconventional mechanisms in the $BiCh_2$-based superconductor family have been proposed by several theoretical and experimental studies [19,20,28-30]. Particularly, the ARPES experiment observed the existence of accidental nodes in nodal *s*-wave symmetry and proposed several possibilities of unconventional pairing mechanisms with competition or cooperation among multiple pairing interactions, such as phonon, charge, and spin fluctuations [20]. Indeed, from neutron diffraction and pair density function analysis, the importance of charge fluctuation to the superconductivity of $La(O,F)BiS_2$ has been proposed [29]. In addition, pairing mechanisms mediated by orbital fluctuation is also possible. Since our isotope effect indicates that phonon is not essential for superconductivity, the mechanisms that are purely electronic or dominated by electronic contribution would drive the superconductivity in $LaO_{0.6}F_{0.4}Bi(S,Se)_2$. Our present results for the Se isotope effect on $T_c$ of $LaO_{0.6}F_{0.4}Bi(S,Se)_2$ should be an important step to clarify the mechanisms for the superconductivity of $BiCh_2$-based layered superconductors.

In conclusion, we have investigated the isotope effect on $T_c$ of $BiCh_2$-based layered superconductor $LaO_{0.6}F_{0.4}Bi(S,Se)_2$ using $^{76}Se$ and $^{80}Se$ isotopes. Comparing the transition temperatures investigated from magnetization and electrical resistivity measurements, we have revealed that the exponent $\alpha_{Se}$ is close to zero. Our results suggest that the pairing in the $LaO_{0.6}F_{0.4}Bi(S,Se)_2$ superconductor is not mediated by phonons, and unconventional superconductivity may emerge in the $BiCh_2$ layer. To completely exclude the phonon-mediated mechanisms, we have to examine the S isotope effect for the same composition ($LaO_{0.6}F_{0.4}Bi(S,Se)_2$) and for a system with a pure $BiS_2$ layer, such as $Nd(O,F)BiS_2$ with a higher $T_c$ of 5 K.

References

1. J. Bardeen, L. N. Cooper, and J. R. Schrieffer, *Phys. Rev.* 108, 1175–1204 (1957).
2. D. G. Hinks *et al.*, *Nature* 335, 419–421 (1988).
3. A. P. Ramirez *et al. Phys. Rev. Lett.* 68, 1058–1060 (1992).
4. S. L. Bud'ko *et al.*, *Phys. Rev. Lett.* 86, 1877–1880 (2001).
5. D. D. Lawrie and J. P. Franck, *Physica C* 245, 159–163 (1995).

6. B. Batlogg *et al.*, *Phys. Rev. Lett.* 58, 2333–2336 (1987).
7. C. C. Tsuei *et al.*, *Phys. Rev. Lett.* 65, 2724–2727 (1990).
8. R. H. Liu *et al.*, *Nature* 459, 64–67 (2009).
9. P. M. Shirage *et al. Phys. Rev. Lett.* 103, 257003(1–4) (2009).
10. Y. Mizuguchi *et al.*, *Phys. Rev. B* 86, 220510(1–5) (2012).
11. Y. Mizuguchi *et al.*, *J. Phys. Soc. Jpn.* 81, 114725(1–5) (2012).
12. Y. Mizuguchi, *J. Phys. Chem. Solids*, 84, 34–48 (2015).
13. Y. Mizuguchi *et al.*, *J. Phys. Soc. Jpn.* 83, 053704(1–4) (2014).
14. H. Usui and K. Kuroki, *Nov. Supercond. Mater.* 1, 50–63 (2015).
15. X. Wan *et al.*, *Phys. Rev. B* 87, 115124(1–6) (2013).
16. S. F. Wu *et al.*, *Phys. Rev. B* 90, 054519(1–5) (2014).
17. G. Lamura *et al.*, *Phys. Rev. B* 88, 180509(1–5) (2013).
18. T. Yamashita *et al.*, *J. Phys. Soc. Jpn.* 85, 073707(1–4) (2016).
19. C. Morice *et al.*, *Phys. Rev. B* 95, 180505(1–6) (2017).
20. Y. Ota *et al.*, *Phys. Rev. Lett.* 118, 167002(1–6) (2017).
21. Y. Mizuguchi *et al.*, *Sci. Rep.* 5, 14968(1–8) (2015).
22. T. Hiroi *et al.*, *J. Phys. Soc. Jpn.* 84, 024723(1–4) (2015).
23. K. Nagasaka *et al.*, *J. Phys. Soc. Jpn.* 86, 074701(1–6) (2017).
24. Y. Mizuguchi *et al.*, *Phys. Chem. Chem. Phys.* 17, 22090–22096 (2015).
25. H. Usui, K. Suzuki and K. Kuroki, Phys. Rev. B 86, 220501(1-5) (2012).
26. F. Izumi and K. Momma, *Solid State Phenom.* 130, 15–20 (2007).
27. K. Momma and F. Izumi, *J. Appl. Crystallogr.* 41, 653–658 (2008).
28. J. Liu *et al.*, *EPL* 106, 67002(p1-p6) (2014).
29. A. Athauda *et al.*, *Phys. Rev. B* 91, 144112(1–6) (2015).
30. K. Suzuki *et al.*, Phys. Rev. B 96, 024513(1-6) (2017).

Acknowledgements

We thank O. Miura and R. Higashinaka for their experimental support and K. Kuroki for his fruitful discussion. This work was partly supported by a Grant-in-Aid for Scientific Research (Nos. 15H05886, 16H04493, 17K19058, and 16K17944) and JST-CREST (No. JPMJCR16Q6), Japan.

Table 1. Information about used isotope, refined crystal structure parameters, and superconducting transition temperatures of $LaO_{0.6}F_{0.4}Bi(S,Se)_2$ samples examined in this study. The atomic coordinates used in the refinements are La(0, 0.5 $z$), O/F(0, 0, 0), Bi(0, 0.5, $z$), Ch1(0, 0.5, $z$), Ch2(0, 0.5, $z$).

| Label | $^{76}$Se–#1 | $^{76}$Se–#2 | $^{80}$Se–#1 | $^{80}$Se–#2 |
|---|---|---|---|---|
| Isotope | $^{76}$Se (99.80%) | $^{76}$Se (99.80%) | $^{80}$Se (99.91%) | $^{80}$Se (99.91%) |
| Space group | $P4/nmm$ | $P4/nmm$ | $P4/nmm$ | $P4/nmm$ |
| $a$ (Å) | 4.13711(5) | 4.13887(5) | 4.13567(4) | 4.13917(4) |
| $c$ (Å) | 13.6022(2) | 13.6031(2) | 13.6014(2) | 13.6333(2) |
| $V$ (Å$^3$) | 232.811(6) | 233.024(6) | 232.638(4) | 233.576(4) |
| $z$ (La) | 0.09603(9) | 0.09639(8) | 0.09638(8) | 0.09629(7) |
| $z$ (Bi) | 0.62834(9) | 0.62853(10) | 0.62862(8) | 0.62865(8) |
| $z$ (Ch1) | 0.3770(2) | 0.3755(2) | 0.3767(2) | 0.3774(2) |
| $z$ (Ch2) | 0.8180(3) | 0.8187(3) | 0.8180(3) | 0.8179(3) |
| Se occupancy at Ch1 | 0.980(7) | 0.958(7) | 0.973(6) | 0.988(5) |
| Se occupancy at Ch2 | 0.121(6) | 0.136(6) | 0.155(6) | 0.149(7) |
| $x$ in $LaO_{0.6}F_{0.4}BiS_{2-x}Se_x$ | 1.101(13) | 1.094(13) | 1.128(12) | 1.137(12) |
| $R_{wp}$ (%) | 9.1 | 9.3 | 8.4 | 7.2 |
| $T_c$ (K)_magnetization | 3.77 | 3.76 | 3.76 | 3.77 |
| $T_{irr}$ (K) _magnetization | 3.74 | 3.74 | 3.74 | 3.74 |
| $T_c$ (K)_resistivity | 3.73 | - | 3.73 | 3.72 |

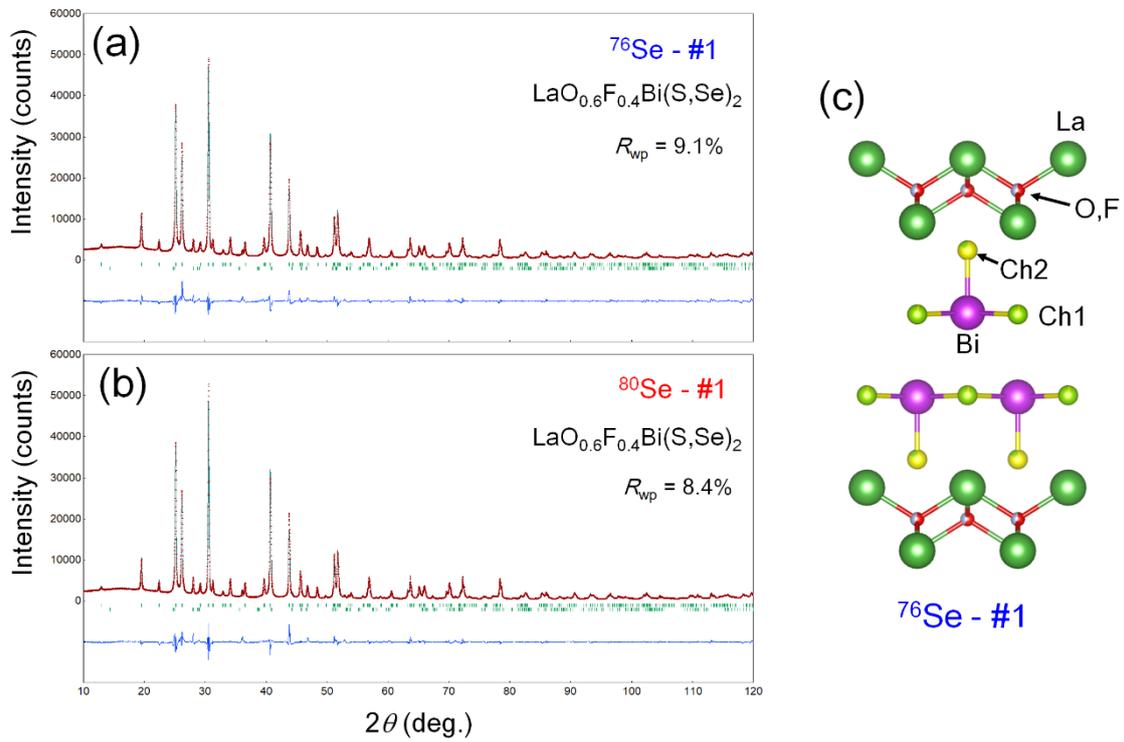

Figure 1. (a,b) X-ray diffraction patterns and Rietveld refinement results for the $^{76}$Se–#1 and $^{80}$Se–#1 samples. The refinements were performed by two-phase analysis with the secondary phase of BiF$_3$ (3%). The blue profiles plotted at the bottom are the differences between observed and calculated patterns. (c) Schematic image of the refined crystal structure for the $^{76}$Se–#1 sample. The Ch1 site is the in-plane chalcogen site.

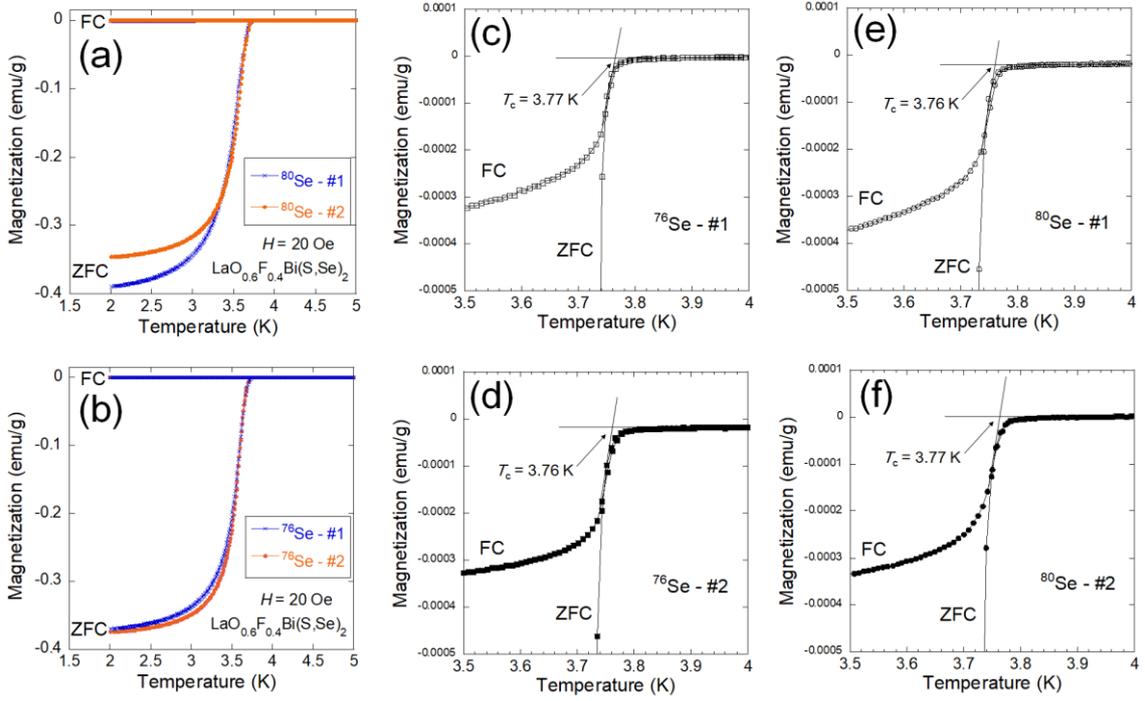

Figure 2. (a) Temperature dependences of magnetization from 2 to 5 K for the $^{76}$Se–#1 and $^{76}$Se–#2 samples. (b) Temperature dependences of magnetization from 2 to 5 K for the $^{80}$Se–#1 and $^{80}$Se–#2 samples. ZFC and FC denote data measured after zero-field cooling and field cooling, respectively. (c–f) Enlarged temperature dependences of magnetization around the superconducting transition. The onset $T_c$ is estimated as 3.77, 3.76, 3.76, and 3.77 K for $^{76}$Se–#1, $^{76}$Se–#2, $^{80}$Se–#1, and $^{80}$Se–#2, respectively. Irreversible temperature $T_{irr}$, defined as the temperature at which the difference between ZFC and FC curves emerges, is also estimated from these plots and listed in Table 1.

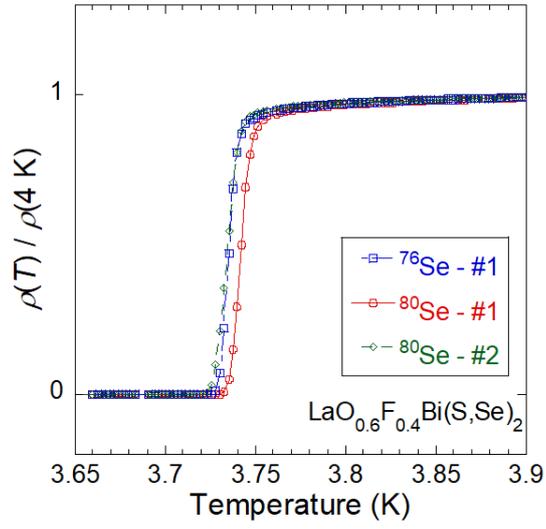

Figure 3. Temperature dependences of normalized electrical resistivity [$\rho(T) / \rho(4\ \text{K})$] for the $^{76}$Se–#1 and $^{80}$Se–#1 samples around the superconducting transition. The onset $T_c$ is estimated as 3.74 and 3.75 K for $^{76}$Se–#1 and $^{80}$Se–#1, respectively. For both samples, zero resistivity is observed at $T_c^{\text{zero}} = 3.73$ K.